# Spin-Lattice Coupling in K$_{0.8}$Fe$_{1.6}$Se$_2$ and KFe$_2$Se$_2$: Inelastic Neutron Scattering and ab-initio Phonon Calculations

R. Mittal[1], M. K. Gupta[1] and S. L. Chaplot[1]

[1]*Solid State Physics Division, Bhabha Atomic Research Centre, Trombay, Mumbai 400 085, India*

M. Zbiri[2], S. Rols[2] and H. Schober[2,3]

[2]*Institut Laue-Langevin, BP 156, 38042 Grenoble Cedex 9, France*

[3]*Université Joseph Fourier, UFR de Physique, 38041, Grenoble Cedex 9, France*

Y. Su[4] and Th. Brueckel[4,5]

[4]*Juelich Centre for Neutron Science JCNS-FRM II, Forschungszentrum Juelich GmbH, Outstation at FRM II, Lichtenbergstr. 1, D-85747 Garching, Germany*

[5]*Juelich Centre for Neutron Science JCNS and Peter Gruenberg Institut PGI, JARA-FIT, Forschungszentrum Juelich GmbH, 52425 Juelich, Germany*

T. Wolf[6]

[6]*Institut fuer Festkoerperphysik, Karlsruhe Institute of Technology, D-76021 Karlsruhe, Germany*

We report measurements of the temperature dependence of phonon densities of states in K$_{0.8}$Fe$_{1.6}$Se$_2$ using inelastic neutron scattering technique. While cooling down to 150 K, a phonon peak splitting around 25 meV is observed and a new peak appears at 31 meV. The measurements support the recent Raman and infra-red measurements indicating a lowering of symmetry of K$_{0.8}$Fe$_{1.6}$Se$_2$ upon cooling below 250 K. Ab-initio phonon calculations have been carried out for K$_{0.8}$Fe$_{1.6}$Se$_2$ and KFe$_2$Se$_2$. The comparison of the phonon spectra as obtained from the magnetic as well as non magnetic calculations show pronounced differences. We show that in the two calculations the energy range of the vibrational contribution from both Fe and Se are quite different. We conclude that Fe magnetism is correlated to the phonon dynamics and it plays an important role in stabilizing the structure of K$_{0.8}$Fe$_{1.6}$Se$_2$ as well as that of KFe$_2$Se$_2$. The calculations highlight the presence of low energy librational modes in K$_{0.8}$Fe$_{1.6}$Se$_2$ as compared to KFe$_2$Se$_2$.

# I. INTRODUCTION

The discovery of superconductivity in La based FeAs ($T_c$=26 K) compounds has stimulated tremendous interest [1-28] in the field of condensed matter physics. So far, the highest Tc of 56 K has been found in $Gd_{1-x}Th_xOFeAs$ [3]. The structural, magnetic, electronic properties of all the compounds have been extensively investigated [4-6] to understand the mechanism of the superconductivity. The long-range antiferromagnetic order is suppressed with doping in the parent compound. The superconductivity in the compounds appears above a certain doping level. In particular for all these compounds strong anomalies have been found in the specific heat, resistivity and magnetic susceptibility in the temperature range of 110 to 180 K. These anomalies are now known to be prerequisite for superconductivity in FeAs compounds.

Adding to the excitement generated by the discovery of iron-based compounds the newly discovered alkali-doped iron selenide compounds [13-16] exhibit several unique characters that are noticeably absent in other iron-based superconductors, such as antiferromagnetically ordered insulating phases and extremely high Neel transition temperatures. Neutron-diffraction studies indicate a very large ordered moment and high magnetic transition temperature for these compounds. The long range magnetic order coexists with superconductivity. The $T_c$ of $K_xFe_{2-y}Se_2$ has been reported [13,14] to be about 31 K under the tetragonal space group I4/m. The magnetic phase disappears [15] above $T_N$ ~559 K followed by the creation of Fe vacancies which exhibit an ordering transition at 578 K. The refinement of the neutron diffraction pattern shows that the compound can now be chemically expressed as $KFe_2Se_2$ with a tetragonal symmetry under space group I4/mmm. Chemical substitution of K by Rb or Cs also yields a superconductor at a similarly high $T_C$ as observed in $Rb_xFe_{2-y}Se_2$ [15] and $Cs_xFe_{2-y}Se_2$ [16]. Further, the local structure of $K_xFe_{2-y}Se_2$ has been measured [12] by temperature dependent polarized extended x-ray absorption fine structure (EXAFS).

Density functional theory (DFT) calculations predict systematically weak electron-phonon coupling [7] with a negligible contribution to the superconductivity mechanism in FeAs systems. The role of the phonons for the mechanism of superconductivity is, however, not yet fully established. However, the electron pairing in FeAs compounds is believed to occur beyond the conventional electron-phonon coupling framework. Theoretically the pairing mechanism is proposed [9, 10] to be mediated by exchange of the antiferromagnetic (AF) spin fluctuations which might lead to an interpenetration of electron and hole pockets at the Fermi surface.

Inelastic neutron scattering has been used to study the resonant spin excitations [17-19] in a number of 122 and 1111 and 11 FeAs based compounds. The phonon measurements as well as ab-inito phonon calculations for FeAs compounds support [25, 27] the idea of a possible coupling between spin and lattice degrees of freedom which affects the phonon dynamics. As far as we know only zone centre phonons of $K_{0.8}Fe_{1.6}Se_2$ (also referred to as the $K_2Fe_4Se_5$ phase) have been studied via Raman and infrared spectroscopies [20-22]. In this context we report results of inelastic neutron scattering measurements of the temperature dependence of phonon spectra over the whole Brillouin zone for $K_{0.8}Fe_{1.6}Se_2$. Ab-initio phonon calculations are carried out in $K_{0.8}Fe_{1.6}Se_2$ and $KFe_2Se_2$ for the sake of interpretation and analysis of the observed phonon spectra, by first lifting and then considering the effect of magnetism. This paper is organized as follows: the details of the experimental technique and lattice dynamical calculations are summarized in section II and section III, respectively. Section IV is dedicated to the presentation and discussion of the results. Conclusions are drawn in section V.

## II. EXPERIMENTAL DETAILS

The inelastic neutron scattering experiments on $K_{0.8}Fe_{1.6}Se_2$ were carried out using the IN4C spectrometers at the Institut Laue Langevin (ILL), France. The spectrometer is based on the time-of-flight technique and is equipped with a large detector bank covering a wide range of about $10^o$ to $110^o$ of scattering angle. Single crystals of $K_xFe_{2-y}Se_2$ were grown by the Bridgman method as previously reported [29]. While we have seen the traces of superconductivity in these crystals with $T_C$= 31 K, the samples are dominated by the insulating $K_{0.8}Fe_{1.6}Se_2$ (also referred to as the $K_2Fe_4Se_5$) phase. A polycrystalline sample of 1 gram of $K_{0.8}Fe_{1.6}Se_2$, grounded from several small single-crystal pieces, was mounted into a cryostat in the transmission mode, at 45° to the incident neutron beam. For these measurements we have used an incident neutron wavelength of 2.4 Å (14.2 meV) in neutron energy gain setup. The measurements were performed at 300 K and 150 K at ambient pressure. In the incoherent one-phonon approximation the measured scattering function $S(Q,E)$, as observed in the neutron experiments, is related [30] to the phonon density of states $g^{(n)}(E)$ as follows:

$$g^{(n)}(E) = A < \frac{e^{2W_k(Q)}}{Q^2} \frac{E}{n(E,T) + \frac{1}{2} \pm \frac{1}{2}} S(Q,E) > \qquad (1)$$

$$g^n(E) = B \sum_k \{\frac{4\pi b_k^2}{m_k}\} g_k(E) \qquad (2)$$

where the + or – signs correspond to energy loss or gain of the neutrons respectively and where $n(E,T) = [\exp(E/k_B T) - 1]^{-1}$. *A* and *B* are normalization constants and $b_k$, $m_k$, and $g_k(E)$ are, respectively, the neutron scattering length, mass, and partial density of states of the $k^{th}$ atom in the unit cell. The quantity between < > represents suitable average over all *Q* values at a given energy. 2*W(Q)* is the Debye-Waller factor. The weighting factors $\frac{4\pi b_k^2}{m_k}$ for various atoms in the units of barns/amu are: K: 0.050; Fe: 0.208 and Se: 0.105. The values of neutron scattering lengths for various atoms can be found from Ref. [31].

## III. COMPUTATIONAL DETAILS

The ab-inito phonon calculations for $K_{0.8}Fe_{1.6}Se_2$ and $KFe_2Se_2$ are performed in the nonmagnetic phase as well as by including the effect of magnetic ordering. At room temperature $K_{0.8}Fe_{1.6}Se_2$ crystallizes in I4/m space group and contains 44 atoms in a unit cell. The ab-initio calculations were done using a primitive unit cell of 22 atoms (TABLE I) with vacancies for K(1) and Fe(1) at 2a and 4d Wyckoff sites. In the absence of a fully stoichiometric structure of $K_{0.8}Fe_{1.6}Se_2$ the structure as given in TABLE I is used for the calculations. As discussed below, the calculated phonon spectra using this structure reproduce correctly the observations. In the nonmagnetic phase a 2×2×1 supercell, containing 176 atoms, is used. The magnetic calculations for $K_{0.8}Fe_{1.6}Se_2$ include the antiferromagnetic ordering as discussed in Ref. [13]. The size of the magnetic cell of $K_{0.8}Fe_{1.6}Se_2$ is the same as that of non magnetic phase. Total energies and inter-atomic forces were calculated for the 22 structures resulting from individual displacements of the five symmetry inequivalent atoms along the three cartesian directions (±x, ±y and ±z).

Upon heating, $K_{0.8}Fe_{1.6}Se_2$ undergoes a non magnetic transition at 559 K followed by an Fe vacancy order transition at 578 K and the compound can now be chemically expressed [13] as $KFe_2Se_2$, under the tetragonal space group I4/mmm with a unit cell containing 10 atoms. Calculations were carried out also in this case using the same supercell (size 2×2×1) approach adopted for $K_{0.8}Fe_{1.6}Se_2$, without and with magnetic ordering. $KFe_2Se_2$ is non magnetic and its structure is similar to $BaFe_2As_2$, which is reported to have [5] C-type magnetic ordering. So for $KFe_2Se_2$ the magnetic ordering was approximated by the C-type ordering. It is worth to note that since $KFe_2Se_2$ is non-magnetic the magnetic calculations are done for the sake of understanding and comparison with $K_{0.8}Fe_{1.6}Se_2$. Total energies and inter-atomic forces were calculated for the 12 structures resulting from individual displacements of the three symmetry inequivalent atoms along the three cartesian directions (±x, ±y and ±z).

The phonon calculations for both the compounds ($K_{0.8}Fe_{1.6}Se_2$ and $KFe_2Se_2$) are carried out in the relaxed configuration. That is the lattice constants as well as atomic coordinates of the atoms are fully optimized in both the magnetic as well as non magnetic cases. Further calculations were also performed with partially relaxed structure, where only the atomic positions were optimized (the lattice parameters were kept at their observed values). Lattice parameters and atomic positions for both the systems are given in Table I.

The Vienna ab initio simulation package (VASP) [32, 33] software is used for the calculations. The projected–augmented wave method and an energy cutoff of plane wave of 600 eV were used. The integrations over the Brillouin zone were sampled on a 8 × 8 × 8 mesh of k-points generated by Monkhorst-pack method [34]. The exchange-correlation contributions were approached within the generalized gradient approximation (GGA) framework and described by the Perdew, Becke and Ernzerhof (PBE) density functional [35,36]. The convergence criteria for the total energy and ionic forces were set to $10^{-8}$ eV and $10^{-5}$ eV Å$^{-1}$, respectively. Phonon spectra for both the compounds were extracted using the PHONON software [37].

## IV. RESULTS AND DISCUSSION

### A. Temperature dependence of phonon spectra

The phonon spectra for $K_{0.8}Fe_{1.6}Se_2$ measured at 315 K and 150 K using the IN4C spectrometer are shown in Figure 1. The high resolution measurements performed with small incident neutron energy of 14.2 meV can be done in the neutron-energy gain mode. At 300 K the peaks in the phonon spectra are located around 10, 25 and 29 meV. When cooling down the peak centered at 25 meV splits into two peaks centered at 23.5 and 26.5 meV in addition to a new peak appearing at about 31 meV. Peak's splitting and the appearance of a new peak indicates a lowering of the symmetry. Our measurements are consistent with recent temperature dependent Raman and infra-red measurements [21] on the stoichiometrically closely similar compound $K_{0.75}Fe_{1.75}Se_2$ where below 250 K new Raman modes appear at about 20.5(165), 24.9 (201) and 26.2(211) meV(cm$^{-1}$) while the new infra red active modes appear at about 12.3(99), 21.2(171) and 30.5(246) meV(cm$^{-1}$). The new modes are non-active Raman (non-active IR) A$_u$(E$_g$) or silent B$_u$ within the space group I4/m. The authors [30] concluded that K$_{0.75}$Fe$_{1.75}$Se$_2$ undergoes a structural phase transition from I4/m to I4, accompanied by the loss of inversion symmetry below 250 K. The Fe-Se stretching around 30 meV are found to shift to slightly higher energies at 150 K as expected from the unit cell volume decrease upon cooling from 300 K to 150 K.

Further, other work using Raman spectroscopy [20] shows that the A$_g$ mode at 8.2 meV (66 cm$^{-1}$) behaves rather anomalously. The intensity of this mode is found to decrease significantly with increase of temperature. Our neutron scattering measurements indicate (Fig. 1) that the intensity of phonon modes around 9 meV exhibits similar behavior. As our measurements were limited to the neutron energy gain mode, no data were collected across the superconducting transition at 31 K. But nevertheless Raman measurements show [20] that the superconducting transition has very little effect on the phonon energies. At the T$_c$, the A$_g$ mode at 180 cm$^{-1}$ exhibits only a jump of approximately 1 cm$^{-1}$.

Next, Figure 2 compares the phonon spectrum of K$_{0.8}$Fe$_{1.6}$Se$_2$ with our previously published data on BaFe$_2$As$_2$ [27]. The energy range of phonon spectra in both the compounds is found to be nearly the same. However, phonon peaks in K$_{0.8}$Fe$_{1.6}$Se$_2$ are broader as compared to those in BaFe$_2$As$_2$. There are three well separated peaks at 20, 25 and 32 meV in the energy range extending from 18 to 35 meV in BaFe$_2$As$_2$, as compared to the broad peaks at 20, 25 and 30 meV in K$_{0.8}$Fe$_{1.6}$Se$_2$. This well supported by EXAFS measurements [12] carried out on K$_{0.8}$Fe$_{1.6}$Se$_2$ at the Fe and Se K-edges and which indicate a large static disorder along the c-axis as well as an Fe site disorder, which may result in broadening of the peaks in the phonon spectra as presently observed. The Fe-Se stretching modes at 30 meV in K$_{0.8}$Fe$_{1.6}$Se$_2$ are found to be shifted to slightly lower energies due to the slightly larger Fe-Se bond length of 2.42 Å in K$_{0.8}$Fe$_{1.6}$Se$_2$ as compared to the Fe-As bond length of 2.4 Å in BaFe$_2$As$_2$ [4].

**B. Effect of magnetic ordering on phonon spectra**

The electronic properties of the FeAs based compounds are known to be sensitively controlled by distortions of the FeAs$_4$ tetrahedra in terms of As-Fe-As bond angle and Fe-As bond length. Earlier it has been shown [28] that the strong interaction between the As ions in FeAs compounds is controlled by the Fe-spin state. The non inclusion of the Fe magnetic moment in the calculations weakens the Fe-As bonding which in turn increases the As-As interactions and decreases their repulsive character, leading to the axial collapse along the c-axis. Considering the iron magnetic moments in the ab-inito calculations leads to a significant change in the electronic structure induced by the correct description of the Fe 3$d$ states near the Fermi level. Consequently, this improves the agreement between the calculated and experimental structural as well as dynamical properties.

The calculated magnetic moment for the equilibrium structure is 2.9 μB and 1.8 μB for $K_{0.8}Fe_{1.6}Se_2$ and $KFe_2Se_2$, respectively. The calculated value of the magnetic moment in $K_{0.8}Fe_{1.6}Se_2$ agrees closely with the experimentally refined one [13] of 3.3 μB at 11 K. $KFe_2Se_2$ is non magnetic and its structure is similar to $BaFe_2As_2$, the calculated value of 1.8 μB is in agreement with the ab-inito calculated value [27] of 1.9 μB for $BaFe_2As_2$.

The neglect of the spin degrees of freedom in the calculations carried out for $K_{0.8}Fe_{1.6}Se_2$ results in a collapse of the c-lattice parameter value from 14.11 Å to 13.39 Å. However by including Fe magnetism, the lattice structure is described correctly (Table I). For the other compound $KFe_2Se_2$, which is reported to be paramagnetic [13] above 580 K, we find that the optimized structure without any magnetic moment at the Fe sites is collapsed along the c-axis, with a value of the axial parameter decreasing from 14.16 Å to 13.42 Å. Interestingly, by including a C-type magnetic ordering on the Fe sites the relaxed structure matches the refined structure without any axial collapse and the estimated c-lattice parameter value (13.70 Å) being close to the experimentally refined one. The detailed comparison between the experimental and calculated structural parameters for both the compounds is given in Table I.

The phonon spectra of $K_{0.8}Fe_{1.6}Se_2$ have been calculated with and without inclusion of the magnetic ordering (Fig. 3). The main difference between the magnetic and non-magnetic calculations is reflected in terms of phonon modes above 20 meV. We find that the magnetic lattice dynamical calculations reproduce correctly the experimental data as compared to the non magnetic case (Fig. 3). The good agreement between the experimental data and magnetic calculations highlights a signature of a magneto-structural correlation, which affects phonon dynamics in $K_{0.8}Fe_{1.6}Se_2$. However a slight difference between estimations and observations is seen and and could originate from the fact that $K_{0.8}Fe_{1.6}Se_2$ [12] has large static disorder along the c-axis as well as large Fe site disorder, which is not included perfectly in the state of the art ab-initio phonon calculations. Further, magnetic phonon calculations were done with fully and partially (by keeping values of the experimental lattice parameters) structural relaxation. The main difference between these two model ab-initio calculations concerns phonon modes above 20 meV. Below this phonon frequency spectra look closely similar. The magnetic calculations based on a partially optimized crystal structure seem however to offer slightly better agreement with the experimental data. Therefore, in the following discussion the magnetic calculations refer to those carried out with partially relaxed structure.

The contribution of the various atomic motions to the phonon spectra can be understood from the calculated partial densities of states (Fig. 4). It is found that vibrations of the K atoms dominate mainly the spectra up to 20 meV in both the magnetic as well as non-magnetic calculations. The atomic vibrations due to Fe and Se span the whole energy range. The inclusion of magnetic ordering in the calculations significantly softens the Fe and Se vibrations as shown in Fig. 4. The Fe-Se stretching modes are above 30 meV.

The differences in the partial densities of states (Fig. 4) in both the calculations can also be understood from the calculated phonon dispersion relations represented in Figure. 5. All the dispersion relations have a large number of slightly dispersive branches up to 20 meV, resulting in a broad peak in the low energy region of the phonon spectra (Fig. 4). The phonon branches in the dispersion relation at 20 meV soften by 5 to 8 meV in the magnetic calculations as compared to the non-magnetic case. The collapse of the c-lattice parameter in the non magnetic calculations from 14.11 Å to 13.39 Å results in a significant change in Fe-Se bond length and in turn overestimates the high energy stretching modes in the non-magnetic calculations.

Neutron diffraction measurements show that $KFe_2Se_2$ stabilizes at ~ 578 K and that it is paramagnetic in nature. The primitive cell of $KFe_2Se_2$ has only 5 atoms in comparison to 22 atoms in $K_{0.8}Fe_{1.6}Se_2$. The increase in symmetry in the high temperature phase results in lowering of a number of phonon branches in the phonon dispersion relation of $KFe_2Se_2$ (Fig. 5). The nearly flat optic branches result in sharp features in the density of states (Fig. 4). Although phonon densities of states (Fig. 1) measurements are carried out only for $K_{0.8}Fe_{1.6}Se_2$, the range of phonon spectra is expected to be the same in both compounds. Here again, as far as the energy range of phonon spectra is concerned, it is found that the calculated phonon spectra including magnetism on the Fe sites compare very well with the experimental data and the other magnetic calculations of $K_{0.8}Fe_{1.6}Se_2$. The calculated (with and without magnetism) partial density of states for $KFe_2Se_2$ (Fig. 4) exhibit the same characteristics as in the case of $K_{0.8}Fe_{1.6}Se_2$. Our calculations clearly support that although $KFe_2Se_2$ is paramagnetic in nature Fe magnetism is always present even at temperatures well above the magnetic ordering temperature and controls the structure, phonon energies, and most probably the superconducting properties of FeSe based compounds.

Recently ab-inito phonon dispersion calculations [11] have been reported for $KFe_2Se_2$. The authors found that phonon branches around 100–150 cm$^{-1}$ (~12-18 meV) in their non magnetic calculations soften by ~ 50 cm$^{-1}$ (~6 meV) when magnetic interactions are considered while calculating the inter-atomic force constants for the dynamical matrix. The high energy phonon modes also show slight softening in the magnetic calculations. These results seem to be inconsistent with our calculations. It is well established that unit cell optimization in the ab-initio calculations without including spin polarization leads to a collapse of the c-lattice parameter which in turn results in an overestimation of the energies of the stretching modes. As discussed above our calculations (which are in agreement with previous phonon calculations [25, 27, 28] on FeAs-based compounds) show that in magnetic calculations phonon modes in the energy range from 12-18 meV soften by about 2 meV. We find that Fe-Se stretching modes around 35-45 meV as calculated in the non-magnetic case soften by about 8 meV when magnetic degrees of freedom are included.

**C. Zone centre phonon modes**

At ambient conditions $K_{0.8}Fe_{1.6}Se_2$ crystallizes in the tetragonal structure I4/m. The group theoretical decomposition of the phonon modes at the zone centre is given by

$$\Gamma = 9A_g + 8B_g + 8E_g + 9A_u + 7B_u + 10E_u$$

Similarly the 15 phonon modes at zone centre within the space group I4/mmm can be classified as

$$\Gamma = A_g + B_g + 2E_g + 3A_u + 3E_u$$

The $A_g$, $B_g$ and $E_g$ modes are Raman active, while $A_u$, $B_u$ and $E_u$ are infrared active. The $B_u$ modes are silent. The calculated zone centre modes of $KFe_2Se_2$ and $K_{0.8}Fe_{1.6}Se_2$ are given in Tables II and III, respectively. Experimental Raman data [21] as available for $K_{0.8}Fe_{1.6}Se_2$ are also presented in Table III. The calculations including the magnetic ordering compare very well with the available experimental Raman data. The high stretching energy modes are found to soften by about 6 to 8 meV, in agreement with the calculated density of states as well as phonon dispersion relations (Figs. 3,4 and 5). The eigenvectors of a few of the selected modes showing the large softening are plotted in Figure. 6.

The $A_g$ mode of $KFe_2Se_2$ (Fig. 6) involves only displacement of Se atoms in the opposite direction along the c-axis, whereas K and Fe atoms are at rest. This mode is found to soften by 6 meV in the magnetic calculation. This clearly shows that neglecting the magnetic ordering at the Fe sites in the calculations has a crucial influence on the binding of the compound and therefore on the phonon frequencies. The c-axis collapses and changes the height (z parameter) of Se atoms and in turn results in an overestimation of the phonon energies in the non-magnetic calculations.

**D. Mean squared displacements of various atoms**

Figure 7 shows the calculated mean squared displacements of the various atoms, expressed as $<(u^2)>$ and arising from all phonons of energy E within the Brillouin zone. This is relevant to the understanding of the averaged nature of phonons in the entire Brillouin zone. The calculated phonon densities of states (Fig. 4) were used for these calculations. The structures of $K_{0.8}Fe_{1.6}Se_2$ and $KFe_2Se_2$ consist of Fe-Se layers, separated by K atoms at c/2. Each of the Fe-Se layers has $FeSe_4$ and $SeFe_4$ units. For $K_{0.8}Fe_{1.6}Se_2$, it is found that K atoms have very large contributions from phonon modes of energies in the range 10-20 meV, indicating a weak interaction between the K atoms and Fe-Se layers. The calculated $<(u^2)>$ for K atoms seems not to be sensitive to whether magnetic ordering is considered or not. The structure of $K_{0.8}Fe_{1.6}Se_2$ has large number of low energy optic modes as compared to $KFe_2Se_2$. The lowest optic mode in $K_{0.8}Fe_{1.6}Se_2$ is at 7 meV as compared to 12 meV in $KFe_2Se_2$. We find that in the case of $K_{0.8}Fe_{1.6}Se_2$, for the low energy modes below 10 meV, Se atoms have larger amplitude in comparison of the Fe atoms. This indicates the presence of low energy libration modes of the $FeSe_4$ tetrahedral units, in addition to a translational motion. We notice that including the magnetic interaction enhances $<(u^2)>$ of the Se atoms and further improves the amplitude of the libration of $FeSe_4$ tetrahedra. For higher energies above 20 meV, $<(u^2)>$ values for Fe and Se atoms are small, indicating the translation motion of $FeSe_4$ and $SeFe_4$ units.

For $KFe_2Se_2$, the calculated $<(u^2)>$ (Fig. 7) shows that for energy range between 10 to 20 meV, K atoms have very large vibrational amplitudes in comparison to those of Fe and Se atoms, indicating the weak interaction between K atoms and Fe-Se layers. The Fe and Se atoms have nearly the same amplitude in the entire energy range. The inclusion of magnetic interaction in the phonon calculations does not impact the calculated amplitudes of $<(u^2)>$ for the low energy modes, which is in contrast to $K_{0.8}Fe_{1.6}Se_2$.

## V. CONCLUSIONS

We have reported detailed measurements of the temperature dependence of the phonon density-of-states of $K_{0.8}Fe_{1.6}Se_2$ using inelastic neutron scattering technique. While cooling down to 150 K, a phonon peak splitting around 25 meV is observed and a new peak appears at 31 meV. This suggests a structural phase transition below 250 K involving a symmetry lowering. The phonon spectra of $K_{0.8}Fe_{1.6}Se_2$ have been analysed based on detailed non-magnetic as well as magnetic ab initio lattice dynamical calculations. Further, calculations are also performed for the non-magnetic $KFe_2Se_2$ under space group I4/mmm. The calculated partial densities of states show that the range of K vibrations does not have any impact upon inclusion of magnetic ordering while vibrations due to Fe and Se atoms soften. The comparison of the experimental and calculated phonon spectra shows that magnetism should be considered in order to describe correctly phonon dynamics. We show that magnetism due to the unpaired electrons of Fe sites does not vanish upon heating above 578 K, and it controls the structure as well as phonon energies of FeSe based compounds suggesting a spin-lattice coupling.

**ACKNOWLEDGEMENT**

T.W. thanks the Deutsche Forschungsgemeinschaft for financial support under the DFG Priority Program 1458.

[1] Y. Kamihara, T. Watanabe, M. Hirano and H. Hosono, J. Am. Chem. Soc. **130**, 3296 (2008).

[2] H. Takahashi, K. Igawa, K. Arii, Y. Kamihara, M. Hirano and H. Hosono, Nature **453**, 376 (2008).

[3] C. Wang, L. Li, S. Chi, Z. Zhu, Z. Ren, Y. Li, Y. Wang, X. Lin, Y. Luo, S. Jiang, X. Xu, G. Cao and Z. Xu, Europhys. Lett. **83**, 67006 (2008).

[4] M. Rotter, M. Tegel and D. Johrendt, Phys. Rev. Lett. **101**, 107006 (2008).

[5] Y. Su, P. Link, A. Schneidewind, Th. Wolf, P. Adelmann, Y. Xiao, M. Meven, R. Mittal, M. Rotter, D. Johrendt, Th. Brueckel and M. Loewenhaupt, Phys. Rev. B **79**, 064504 (2009).

[6] Y. Xiao, Y. Su, R. Mittal, T. Chatterji, T. Hansen, C. M. N. Kumar, S. Matsuishi, H. Hosono and Th. Brueckel, Phys. Rev. B **79**, 060504(R) (2009).

[7] L. Boeri, O. V. Dolgov and A. A. Golubov, Phys. Rev. Lett **101**, 026403 (2008).

[8] M. Le Tacon, T. R. Forrest, Ch. Rüegg, A. Bosak, A. C. Walters, R. Mittal, H. M. Rønnow, N. D. Zhigadlo, S. Katrych, J. Karpinski, J. P. Hill, M. Krisch, D.F. McMorrow, Phys. Rev. B **80**, 220504(R) (2009).

[9] I. I. Mazin, D. J. Singh, M. D. Johannes and M. H. Du, Phys. Rev. Lett. **101** 057003 (2008).

[10] K.Kuroki, S. Onari, R. Arita, H. Usui, Y. Tanaka, H. Kontani, and H. Aoki, Phys. Rev. Lett. **101** 087004 (2008).

[11] T. Bazhirov and M. L. Cohen, Phys. Rev. B **86**, 134517 (2012).

[12] A. Iadecola, B. Joseph, L. Simonelli, A. Puri, Y. Mizuguchi, H. Takeya, Y. Takano and N. L. Saini, J. Phys.: Condens. Matter **24**, 115701(2012).

[13] W. Bao, Q.-Z. Huang, G.-F. Chen, M. A. Green, D.-M. Wang, J.-B. He, and Y.-M. Qiu, Chin. Phys. Lett. **28**, 086104 (2011).

[14] J. Guo, S. Jin, G. Wang, S. Wang, K. Zhu, T. Zhou, M. He and X. Chen, Phys. Rev. B **82**, 180520(R) (2010). G. Friemel, W. P. Liu, E. A. Goremychkin, Y. Liu, J. T. Park, O. Sobolev, C. T. Lin, B. Keimer and D. S. Inosov, Euro Phys. Lett., **99**, 67004 (2012).

[15] C. H. Li, B. Shen, F. Han, X. Y. Zhu and H. H. Wen Phys. Rev. B **83** 184521 (2011).

[16] A. Krzton-Maziopa, Z. Shermadini, E. Pomjakushina, V. Pomjakushin, M. Bendele, A. Amato, R. Khasanov, H. Luetkens, K. Conder, J. Phys.: Cond. Mat. **23**, 052203 (2011).

[17] A. D. Christianson, E. A. Goremychkin, R. Osborn, S. Rosenkranz, M. D. Lumsden, C. D. Malliakas, L. S. Todorov, H. Claus, D. Y. Chung, M. G. Kanatzidis, R. I. Bewley and T. Guidi, Nature **456**, 930 (2008).

[18] C. de la Cruz, Q. Huang, J. W. Lynn, J. Li, W. Ratcliff II, J. L. Zarestky, H. A. Mook, G. F. Chen, J. L. Luo, N. L. Wang, and P. Dai, Nature **453**, 899 (2008).

[19] W. Bao, Y. Qiu, Q. Huang, M. A. Green, P. Zajdel, M. R. Fitzsimmons, M. Zhernenkov, S. Chang, Minghu Fang, B. Qian, E. K. Vehstedt, Jinhu Yang, H. M. Pham, L. Spinu, and Z. Q. Mao, Phys. Rev. Lett. **102,** 247001 (2011).

[20] A. M. Zhang, K. Liu, J. H. Xiao, J. B. He, D. M. Wang, G. F. Chen, B. Normand and Q. M. Zhang, Phys. Rev. B **85**, 024518 (2012).

[21] A. Ignatov, A. Kumar, P. Lubik, R. H. Yuan, W. T. Guo, N. L. Wang, K. Rabe and G. Blumberg, arXiv: 1209.5718.

[22] C. C. Homes, Z. J. Xu, J. S. Wen, and G. D. Gu, Phys. Rev. B **85**, 180510(R) (2012).

[23] R. Mittal, Y. Su, S. Rols, T. Chatterji, S. L. Chaplot, H. Schober, M. Rotter, D. Johrendt and Th. Brueckel, Phys. Rev. B **78**, 104514 (2008).

[24] R. Mittal, Y. Su, S. Rols, M. Tegel, S. L. Chaplot, H. Schober, T. Chatterji, D. Johrendt and Th. Brueckel, Phys. Rev. B **78**, 224518 (2008).

[25] M. Zbiri, R. Mittal, S. Rols, Y. Su, Y. Xiao, H. Schober, S. L. Chaplot, M. R. Johnson, T. Chatterji, Y. Inoue, S. Matsuishi, H. Hosono, Th. Brueckel, J. Phys.: Condens. Matter **22**, 315701 (2010).

[26] R. Mittal, L. Pintschovius, D. Lamago, R. Heid, K. -P. Bohnen, D. Reznik, S. L. Chaplot, Y. Su, N. Kumar, S. K. Dhar, A. Thamizhavel and Th. Brueckel, Phys. Rev. Lett. **102**, 217001 (2009).

[27] M. Zbiri, H. Schober, M. R. Johnson, S. Rols, R. Mittal, Y. Su, M. Rotter and D. Johrendt, Phys. Rev. B **79**, 064511 (2009).

[28] T. Yildirim, Phys. Rev. Lett. **102**, 037003 (2009).

[29] S. Landsgesell, D. Abou-Ras, T. Wolf, D. Alber and K. Prokes, Phys. Rev. B **86**, 224502 (2012).

[30] D. L. Price and K. Skold, in *Neutron Scattering*, edited by K. Skold and D. L. Price (Academic Press, Orlando, 1986), Vol. A; J. M. Carpenter and D. L. Price, Phys. Rev. Lett. **54**, 441 (1985); S. Rols, H. Jobic and H. Schober, C. R. de Physique **8**, 777 (2007).

[31] www.ncnr.nist.gov; V. F. Sears, Neutron News **3**, 29 (1992); A. -J. Dianoux and G. Lander (Eds.), *Neutron Data Booklet, Institut Laue-Langevin*, Grenoble, France (2002).

[32] G. Kresse and J. Furthmüller, Comput. Mater. Sci. **6**, 15 (1996).

[33] G. Kresse and D. Joubert, Phys. Rev. B **59**, 1758 (1999).

[34] H. J. Monkhorst and J. D. Pack, Phys. Rev. B **13**, 5188 (1976)

[35] J. P. Perdew, K. Burke, and M. Ernzerhof, Phys. Rev. Lett. **77**, 3865 (1996).

[36] J. P. Perdew, K. Burke, and M. Ernzerhof, Phys. Rev. Lett. **78**, 1396 (1997).

[37] K. Parlinksi, Software phonon (2003).

TABLE I. Comparison between the experimental [13] and calculated structural parameters for KFe$_2$Se$_2$ (space group I4/mmm) and K$_{0.8}$Fe$_{1.6}$Se$_2$ (space group I4/m). The calculations are carried out at 0 K, while the experimental data for K$_{0.8}$Fe$_{1.6}$Se$_2$ and KFe$_2$Se$_2$ are given at 295 K and 580 K, respectively. For KFe$_2$Se$_2$ (space group I4/mmm) the K, Fe and Se atoms are located at (0, 0, 0), (0, 0.50, 0.25) and (0, 0, z) respectively and their symmetry equivalent positions. The experimental structure of K$_{0.8}$Fe$_{1.6}$Se$_2$ (space group I4/m) consists [13] of atoms at: K1, 2$a$(0,0,0); K2, 8$h$($x$, $y$, 0), Fe1, 4$d$(0,1/2,1/4); Fe2, 16$i$($x$, $y$, $z$); Se1, 4$e$ (1/2, 1/2, $z$) and Se2, 16$i$ ($x$, $y$, $z$) with site occupancies of 1.06, 0.80, 0.059, 1.020, 1 and 1 respectively. Whereas the ab-initio calculations were done with vacancies at K and Fe at 2a and 4d Wyckoff sites and atomic positions: K, 8$h$($x$, $y$, 0); Fe, 16$i$ ($x$, $y$, $z$); Se1, 4$e$ (1/2, 1/2, $z$) and Se2, 16$i$ ($x$, $y$, $z$) respectively.

| | | KFe$_2$Se$_2$ (space group I4/mmm) | | |
| --- | --- | --- | --- | --- |
| | | Experimental | Fully relaxed Non-magnetic | Fully relaxed magnetic |
| | $a$ (Å) | 3.94502 | 3.8382 | 3.8707 |
| | $c$ (Å) | 14.1619 | 13.4160 | 13.6984 |
| | Z | 0.35444 | 0.3473 | 0.35047 |

| | | | K$_{0.8}$Fe$_{1.6}$Se$_2$ (space group I4/m) | | |
| --- | --- | --- | --- | --- | --- |
| | | Experimental | Fully relaxed Non-magnetic | Partially relaxed magnetic | Fully relaxed magnetic |
| | $a$ (Å) | 8.7308 | 8.6049 | 8.7308 | 8.6764 |
| | $c$ (Å) | 14.1128 | 13.3871 | 14.1128 | 14.4001 |
| K | $x$ | 0.403 | 0.3705 | 0.3766 | 0.3736 |
| | $y$ | 0.180 | 0.1903 | 0.1856 | 0.1855 |
| Fe | $x$ | 0.1984 | 0.1941 | 0.1936 | 0.1940 |
| | $y$ | 0.0918 | 0.1048 | 0.0942 | 0.0942 |
| | $z$ | 0.2515 | 0.2495 | 0.2485 | 0.2484 |
| Se1 | $z$ | 0.1351 | 0.1500 | 0.1339 | 0.1855 |
| Se2 | $x$ | 0.1147 | 0.0958 | 0.1081 | 0.1081 |
| | $y$ | 0.3000 | 0.3055 | 0.3025 | 0.3028 |
| | $z$ | 0.1462 | 0.1604 | 0.1504 | 0.1512 |

TABLE II. The calculated zone centre optic phonon modes (cm$^{-1}$) for KFe$_2$Se$_2$ (space group I4/mmm) (1 meV=8.0585 cm$^{-1}$).

|     | Fully relaxed Non-magnetic | Fully relaxed Magnetic |
|-----|---|---|
| A$_g$ | 224 | 174 |
| B$_g$ | 225 | 188 |
| E$_g$ | 137 | 120 |
|     | 298 | 258 |
| A$_u$ | 117 | 101 |
|     | 292 | 268 |
| E$_u$ | 100 | 96 |
|     | 277 | 248 |

TABLE III. The experimental [21] and calculated zone centre optic phonon modes (cm$^{-1}$) for K$_{0.8}$Fe$_{1.6}$Se$_2$ (space group I4/m) (1 meV=8.0585 cm$^{-1}$).

|     | Experimental | Fully relaxed Non-magnetic | Fully relaxed Magnetic | Partially relaxed Magnetic |
|-----|---|---|---|---|
| A$_g$ | 67.6 | 71 | 65 | 69 |
|     | 81.0 | 85 | 78 | 83 |
|     | 111.8 | 110 | 96.2 | 96 |
|     | 124.8 | 127 | 115 | 116 |
|     | 135.9 | 158 | 133 | 138 |
|     | 182.5 | 209 | 175 | 178 |
|     | 205.3 | 267 | 204 | 213 |
|     | 242.3 | 284 | 229 | 242 |
|     | 267.9 | 317 | 256 | 267 |
| B$_g$ | 63.1 | 64 | 56 | 59 |
|     |  | 83 | 78 | 73 |
|     | 100.6 | 114 | 100 | 108 |
|     | 103.3 | 137 | 127 | 129 |
|     | 143.6 | 188 | 131 | 137 |
|     | 195.3 | 237 | 194 | 197 |
|     | 216.1 | 269 | 209 | 220 |
|     | 277.1 | 324 | 257 | 273 |
| E$_g$ |  | 79 | 64 | 67 |
|     |  | 100 | 77 | 83 |
|     |  | 115 | 101 | 101 |
|     |  | 142 | 112 | 118 |
|     |  | 207 | 153 | 155 |
|     |  | 240 | 199 | 210 |
|     |  | 283 | 212 | 222 |

|     |     |     |     |
| --- | --- | --- | --- |
|     | 324 | 246 | 262 |
| $A_u$ | 71  | 63  | 68  |
|     | 115 | 100 | 107 |
|     | 118 | 108 | 111 |
|     | 210 | 167 | 165 |
|     | 252 | 205 | 218 |
|     | 301 | 237 | 251 |
|     | 307 | 265 | 273 |
| $B_u$ | 84  | 61  | 69  |
|     | 104 | 78  | 80  |
|     | 121 | 96  | 103 |
|     | 145 | 119 | 122 |
|     | 216 | 164 | 171 |
|     | 280 | 230 | 242 |
|     | 317 | 261 | 274 |
| $E_u$ | 64  | 62  | 69  |
|     | 87  | 82  | 82  |
|     | 108 | 95  | 98  |
|     | 131 | 111 | 111 |
|     | 187 | 141 | 144 |
|     | 264 | 205 | 214 |
|     | 277 | 228 | 239 |
|     | 320 | 261 | 273 |

FIG. 1 (Color online) The experimental phonon spectra for $K_{0.8}Fe_{1.6}Se_2$ at 150 K and 300 K.

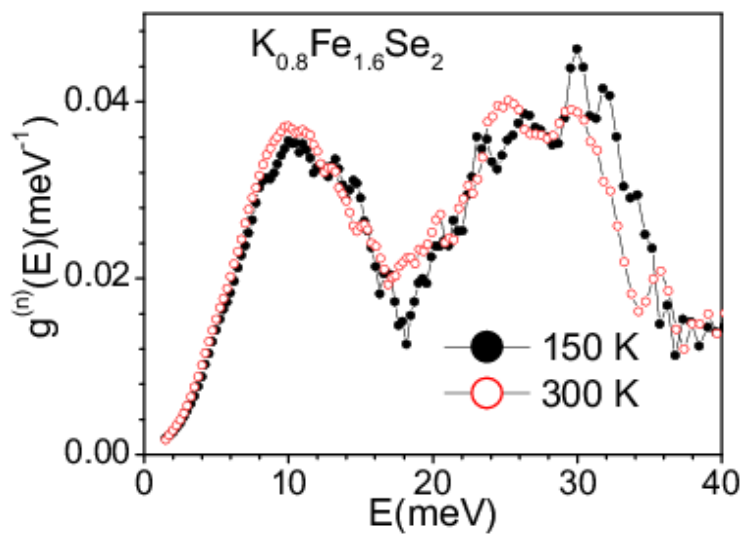

FIG. 2. The experimental phonon spectra for $K_{0.8}Fe_{1.6}Se_2$ and $BaFe_2As_2$ [27] at 300 K.

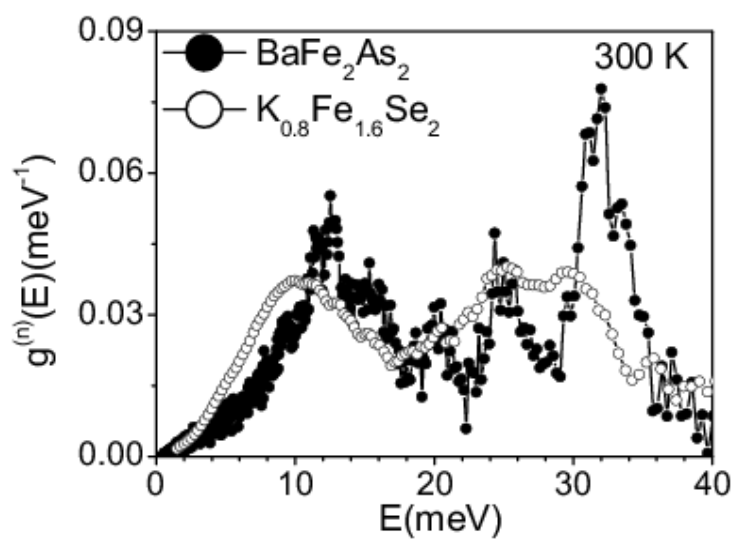

FIG. 3. (Color online) The experimental and calculated phonon spectra for $K_{0.8}Fe_{1.6}Se_2$ (space group I4/m) and $KFe_2Se_2$ (space group I4/mmm). The solid and open symbols correspond to the experimental data at 150 K and 300 K, respectively. "FM", "FNM" and "PM" refer to fully relaxed magnetic, fully relaxed non-magnetic and partially relaxed magnetic calculations.

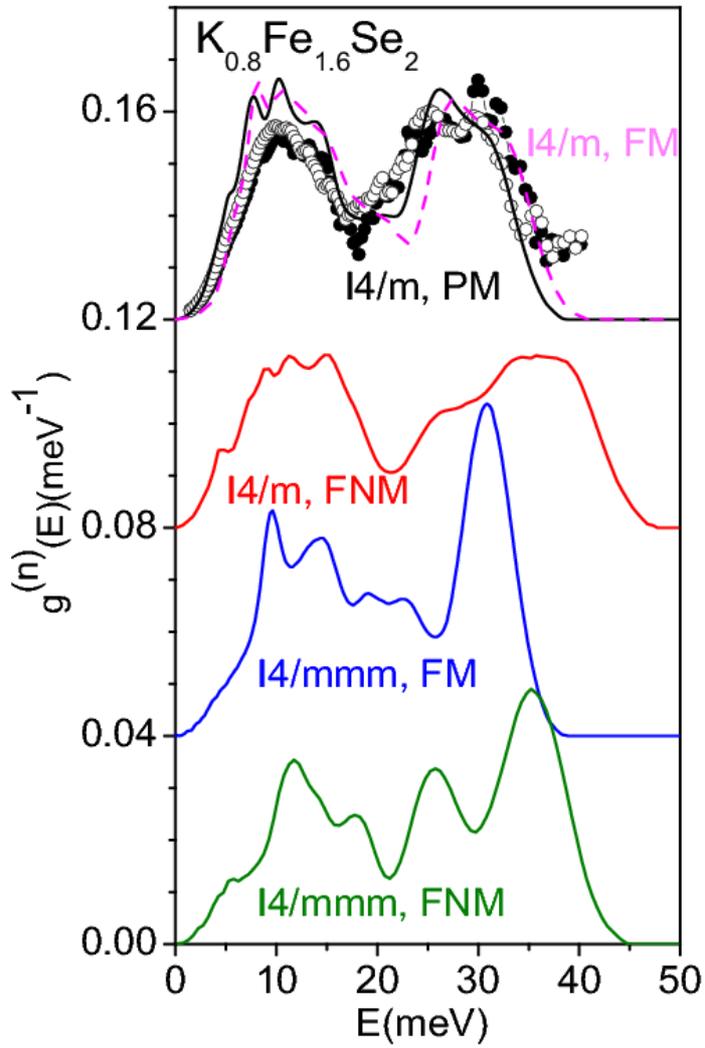

FIG. 4. (Color online) The calculated partial density of states for various atoms and the total one phonon density of states for $K_{0.8}Fe_{1.6}Se_2$ (space group I4/m) and $KFe_2Se_2$ (space group I4/mmm). The solid and dashed lines correspond to magnetic and non-magnetic calculations, respectively. For $K_{0.8}Fe_{1.6}Se_2$, we have shown partially relaxed magnetic phonon calculations, while for $KFe_2Se_2$ we have carried out fully relaxed magnetic calculations.

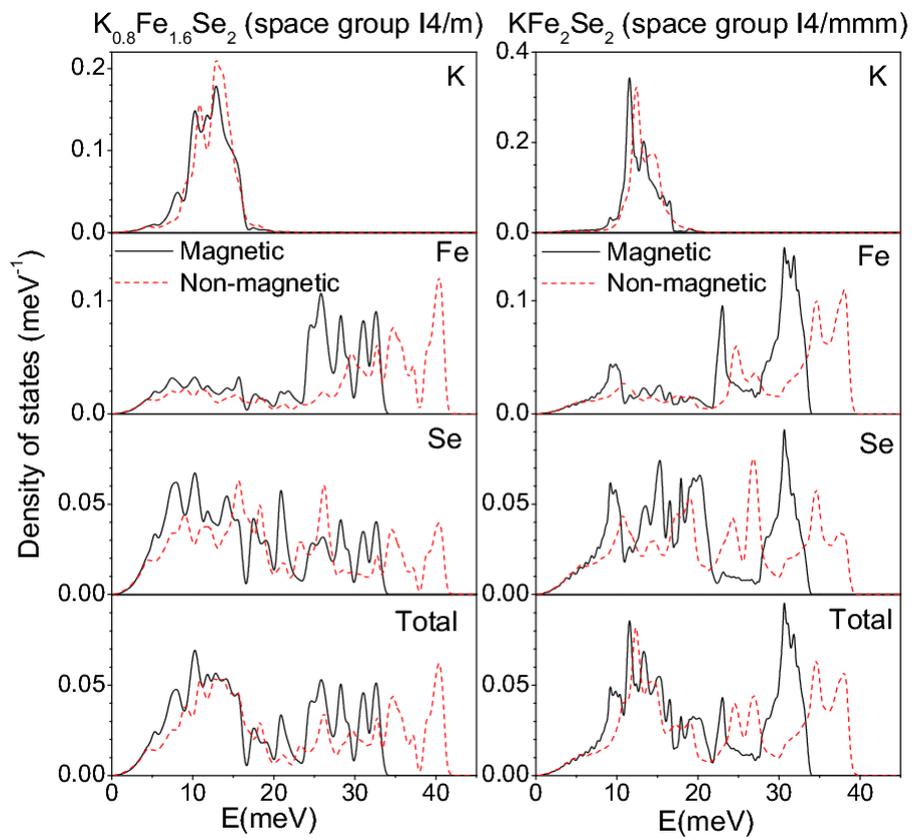

FIG. 5. (Color online) The calculated phonon dispersion curves for $K_{0.8}Fe_{1.6}Se_2$ (space group I4/m) and $KFe_2Se_2$ (space group I4/mmm). The solid and dashed lines correspond to magnetic and non-magnetic calculations, respectively. For $K_{0.8}Fe_{1.6}Se_2$, we have shown partially relaxed magnetic phonon calculations, while for $KFe_2Se_2$ we have only carried out fully relaxed magnetic calculations.

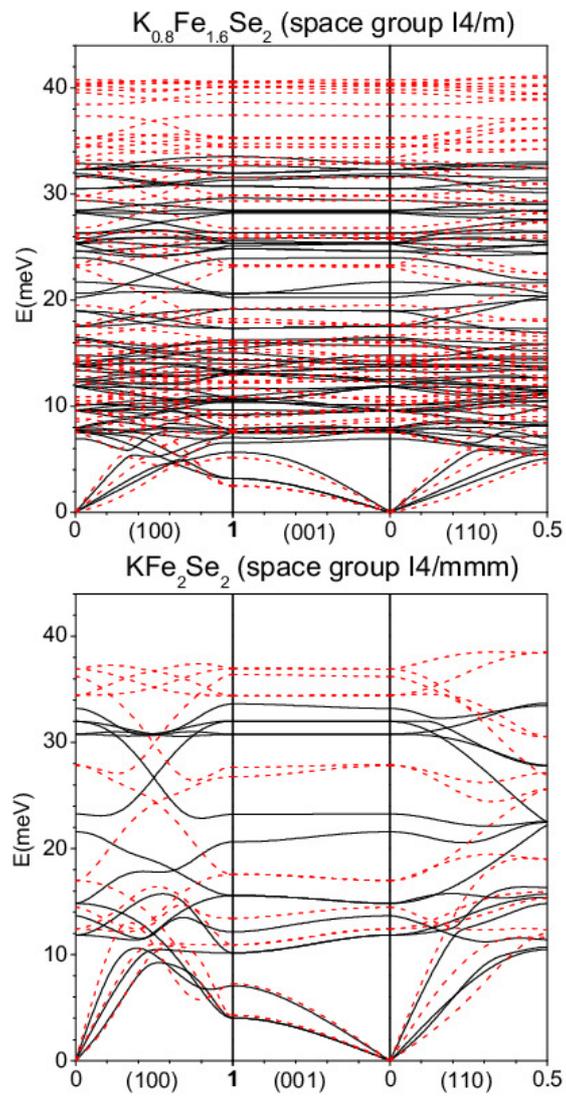

FIG. 6. (Color online) Polarization vectors of selected zone centre modes in KFe$_2$Se$_2$ (space group I4/mmm) and K$_{0.8}$Fe$_{1.6}$Se$_2$ (space group I4/m) showing exceptionally high softening on inclusion of magnetic ordering in the phonon calculation. For each mode the mode assignment and energy is indicated. The lengths of arrows are related to the displacements of the atoms. The absence of an arrow on an atom indicates that the atom is at rest. The numbers after the mode assignments give the phonon energies as calculated from partially relaxed magnetic and fully relaxed magnetic calculations for K$_{0.8}$Fe$_{1.6}$Se$_2$ and KFe$_2$Se$_2$, respectively. The z-axis is vertical, while the x and y-axes are in the horizontal plane. Key: K, blue spheres; Fe, red spheres; Se, yellow spheres (1 meV=8.0585 cm$^{-1}$).

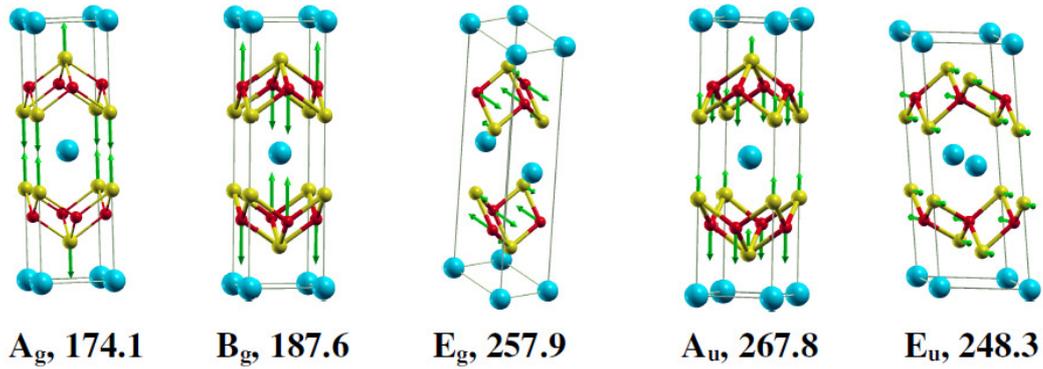

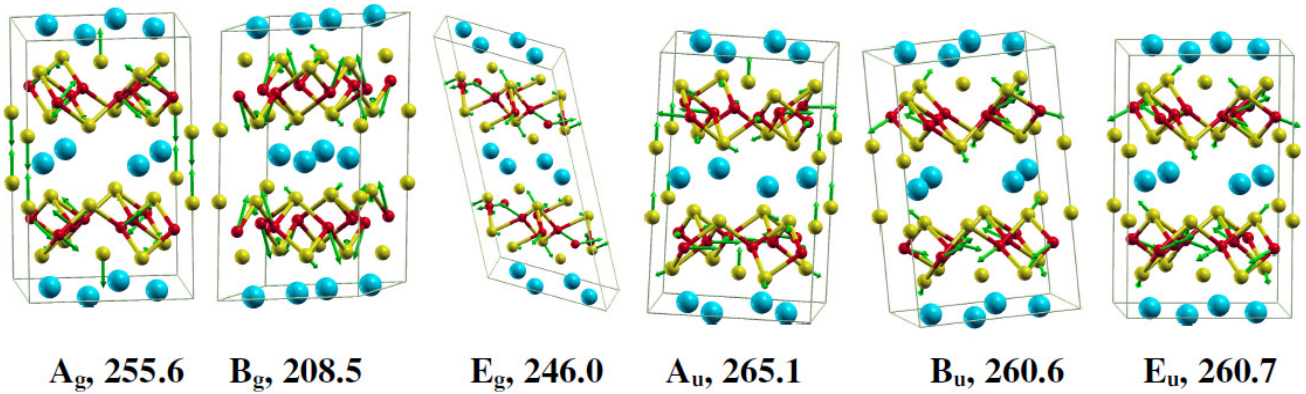

FIG. 7. (Color online) The calculated contribution to the mean squared amplitude of the various atoms arising from phonons of energy E at $T=300$ K and 580 K in $K_{0.8}Fe_{1.6}Se_2$ (space group I4/m) and $KFe_2Se_2$ (space group I4/mmm). For $K_{0.8}Fe_{1.6}Se_2$, we have shown partially relaxed magnetic phonon calculations, while for $KFe_2Se_2$ we have only carried out fully relaxed magnetic calculations.

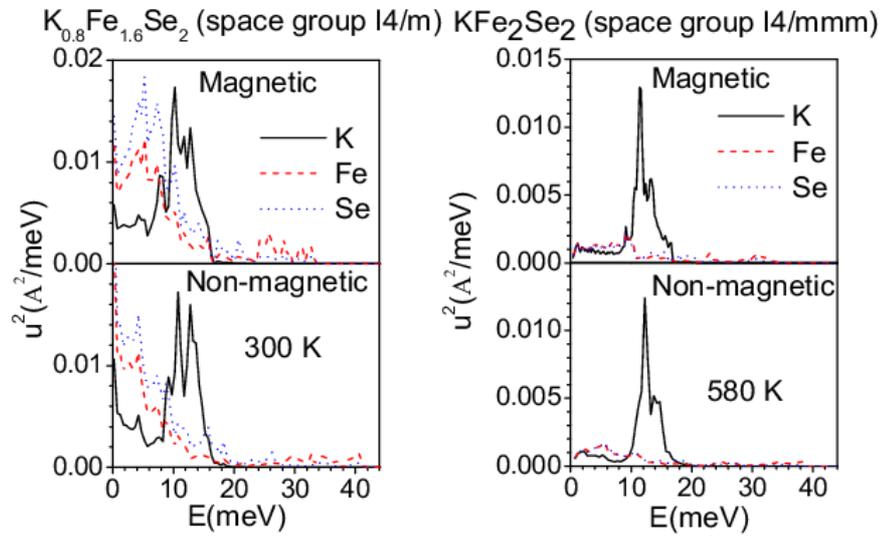